\documentclass[fleqn,10pt]{wlscirep}
\usepackage[utf8]{inputenc}
\usepackage[T1]{fontenc}
\usepackage{lineno}
\usepackage{float}
\usepackage{tabularx}
\usepackage{booktabs}
\usepackage{subcaption}
\usepackage{ltablex}
\usepackage{fvextra}

\title{Light-Adapted Electroretinogram and Oscillatory Potentials (LEOPs) Dataset for Autism Spectrum Disorder and Typically Developing Individuals}

\author[1]{Paul A. Constable}
\author[2,3]{Dorothy A. Thompson}
\author[4]{Irene O. Lee}
\author[1]{Lynne Loh}
\author[5]{Aleksei Zhdanov}
\author[5]{Mikhail Kulyabin}
\author[6]{Andreas Maier}

\affil[1]{Flinders University, College of Nursing and Health Sciences, Caring Futures Institute, Adelaide, Australia}
\affil[2]{The Tony Kriss Visual Electrophysiology Unit, Clinical and Academic Department of Ophthalmology, Great Ormond Street Hospital for Children NHS Trust, London, United Kingdom}
\affil[3]{UCL Great Ormond Street Institute of Child Health, University College London, London, United Kingdom}
\affil[4]{Behavioural and Brain Sciences Unit, Population Policy and Practice Programme, UCL Great Ormond Street Institute of Child Health, University College London, London, United Kingdom}
\affil[5]{Visiomed.AI, Moscow, Russia}
\affil[6]{Pattern Recognition Lab, Friedrich-Alexander-Universität Erlangen-Nürnberg, 91058 Erlangen, Germany}

\begin{abstract}
The LEOPs (Light-ERG-Oscillatory Potentials) dataset provides light-adapted (LA) electroretinogram (ERG) and Oscillatory Potentials (OPs) waveforms for typically developing Control, Autism Spectrum Disorder (ASD) and ASD + Attention Deficit Hyperactivity Disorder (ADHD) childhood and adolescent populations. The ERGs were recorded in the Right And Left eyes with skin electrodes using the handheld RETeval device at two sites in Australia and the United Kingdom. The LEOPs dataset includes 5309 single flash ERG and 4434 OPs waveforms as well as images selected from each participant showing the position of the skin electrode. The LEOPs dataset is constructed from recordings using a 9 step randomized flash series from $-0.37$ to $1.20$~$Td.s$, a 2 step at 113 and 446 $Td.s$ flash strengths (2500 Control, 1730 ASD and 451 ASD + ADHD samples), as well as the $85$~$Td.s$ (Light Adapted 3 $cd.s.m^{-2}$ (LA3)) equivalent International Society of Clinical Electrophysiology of Vision (ISCEV) Standard flash with 435 Control, 176 ASD and 37 ASD + ADHD waveform samples. Code for the stimulus is provided along with participant demographics, date and time of testing, and where available diagnostic scores for the ASD and ASD + ADHD groups, alongside iris color, electrode position with image files and time domain values for the ERG and summed values for the OPs. The repository contains excel file, exported JSON files on the patient level that are more suitable for machine learning tasks, images of electrode position for each recording and the protocol files for use with the RETeval.
\end{abstract}
\begin{document}

\flushbottom
\maketitle
        
\thispagestyle{empty}

\section*{Background \& Summary}
The LEOPs dataset \cite{constable2026LEOPs} provides light adapted (LA) electroretinogram (ERG) and Oscillatory Potentials (OPs) averaged  waveforms for the  (right and left eyes with replicates) in children and adolescents diagnosed with autism spectrum disorder (ASD), ASD + Attention Deficit Hyperactivity Disorder (ADHD) and typically developing Control individuals. The full-field ERG is a biological signal recorded from the retina in response to a brief flash of light. The resulting waveform is generated from photoreceptors, bipolar, glial, horizontal, amacrine and ganglion cells \cite{robson2003rod, wachtmeister1998oscillatory, viswanathan2001photopic, friedburg2004contribution}. The International Society for Clinical Electrophysiology of Vision (ISCEV) recommends a series of standard ERGs \cite{robson2022iscev} and supports the availability of normative databases to be publicly available to integrate into a harmonized reference dataset library \cite{hamilton2021clinical}. The LEOPs dataset  provides ISCEV standard light-adapted  3 $cd.s.m^{-2}$ (LA3) ERGs for ASD and Control groups as well as a nine-flash strength sequence that describes the light-adapted stimulus response function known as the photopic hill \cite{wali1992photopic}. In addition, where available the OPs waveforms are provided to explore the high frequency components of the ERG waveform \cite{gauvin2016assessing}. 

The LEOPs dataset provides potential to use the Control waveforms as part of a normative database or reference comparison group. The ASD group dataset could be used in conjunction with other biological or psychophysical biomarkers in studies exploring the classification of ASD from other groups using AI \cite{patel2025ai}. The OPs waveforms provide an additional resource with which to explore the changes in the OPs over a wide 9 step flash series. In addition, the LEOPs dataset includes the standard LA3 response that is part of the ISCEV recommended flash series \cite{robson2003rod}. Importantly, images of the skin electrode position are provided that are linked to each subject and recording  so that future analyses could be made to ascertain the impact of skin electrode position of the recorded signal \cite{hobby2018effect}. Iris color data is also included in the LEOPs dataset to compensate for the impact of iris color/ pigment on the ERG amplitude \cite{al2010light}. Time of day and date for each recording is also included to potentially study circadian variations in the ERG waveform \cite{nozaki1983circadian,hankins1998diurnal}. The LEOPs dataset as recorded using the RETeval (LKC Technologies, Germantown, MD, USA) with skin electrodes. Therefore, the LEOPs dataset would provide a comparative dataset for ERGs recorded with other electrode types such as thread, gold foil or contact lens \cite{dubois2025validating}. 

Prior publications based on these data included attempts to classify ASD from Control and ADHD groups using machine learning using spectral domain analysis including Discrete Wavelet Transform, Variable Frequency Complex Demodulation and Continuous Wavelet Transform that provide moderate to high classification for neurodevelopmental groups with > 80\% accuracy \cite{manjur2025detecting, manjur2022detecting, posada2023autism, constable2024spectral}. Time series-based analysis to identify key parts of the ERG waveform to inform classification models \cite{chistiakov2025time}. The application of a gated multi-layer perceptron analysis \cite{kulyabin2024attention} yielded a classification accuracy of 89.7\%. Finally short-time Fourier transforms have been used to generate features for deep learning to classify these groups achieving an AUC of 0.89 \cite{albasu2024electroretinogram}.

The averaged waveforms have been used to generate synthetic ERG waveforms to bolster the training and testing datasets for ML models as well as provide an increase in sample size to balance unbalanced datasets \cite{kulyabin2025synthetic, kulyabin2024generating}. Further analyses have incorporated aspects of Functional Data Analysis and Bayesian statistics to model the rising portion of the b-wave to discriminate changes in the dynamic profile of the ERG waveform between the ASD and Control groups \cite{brabec2026contour, brabec2023group, brabec2025remodeling}. The waveforms have also been used as an additional marker for psychophysical markers in ASD and typically developing children \cite{lee2025global}. Whilst the LEOPs dataset comprises ASD, ASD + ADHD, and Control groups further applications could be used to differentiate between the wider neurodevelopmental spectrum including ADHD \cite{lee2022electroretinogram, constable2022discrete, manjur2023spectral, dubois2023evaluation} and drug interactions with the retina \cite{huang2024retinal, constable2025brief,ergun2026retinal}. Whilst the main features of the LEOPs dataset demonstrate group differences have been observed in the b-wave of the ERG, studies focusing on the later Photopic Negative Response (PhNR) that is driven by ganglion cells could also utilize these data for retinal and neurodevelopmental disorders \cite{viswanathan2001photopic, dubois2025validating, constable2021photopic}. The available protocol files have also been used in one prior study investigating the ERG in adults with ASD \cite{friedel2022electroretinography} and could be used for related neurological disorders in which the LA-ERG that form the photopic hill are atypical such as Parkinson's Disease, schizophrenia and Bipolar disorder \cite{constable2023retinal,soto2025enhancing, hebert2020electroretinogram, ergun2026retinal}.

The LEOPs dataset also includes the OPs waveforms that are the focus of novel analytical methods including the Hilbert Transform \cite{gauthier2025hilbert} and will provide data with which to explore ways of modeling these waveforms. The photopic hill data series may also be beneficial to expanding the mathematical modeling of this stimulus-response function \cite{hamilton2007luminance,mcculloch2019iscev}. The 9 step series provides a resource with which to model the a-wave kinetics across an extended flash series \cite{mahroo2004recovery} and the PhNR that also has different modalities in which to assess \cite{tzekov2026area, prencipe2020photopic}. The LEOPs dataset will provide additional resources for the classification of neurodevelopmental disorders that use functional and/or structural biomarkers \cite{constable2023retinal, silverstein2020schizophrenia, constable2016full, ritvo1988electroretinograms, hosseinzadeh2025optical} that span Visual Transformers \cite{kulyabin2023enhancing}, AI modeling of pediatric ERGs \cite{kulyabin2023optimal} to support the future development of AI driven tools for mental health \cite{peredo2022developing}. 
	
Related datasets include in time series values for ASD and Control groups and ISCEV standard ERGs recorded with the RETeval but no raw waveforms \cite{constable2026dataset}. Pattern ERG waveforms have been made publicly available for assessment of macular function \cite{fernandez2024comprehensive}. Normative ranges but no waveforms for skin electrodes in a pediatric population are also published \cite{wang2023full}. Reference ranges in a Chinese childhood population for the ISCEV standard series using the RETeval is also available but without waveform, data \cite{chan2025reference}. Reference limits for the full field dark and light adapted ERGs in 407 participants with three electrode types across two sites is available \cite{baker2025iscev}. Synthetically generated ERG waveforms from four flash strengths of ASD and Control groups is available at IEEE DataPort \cite{npv7-8063-24}. In contrast to current publicly available datasets the LEOPs dataset provides averaged waveforms in JSON and excel files for both ERG and OPs, collected at two sites with detailed demographic data for Control, ASD and ASD + ADHD populations. Psychological test data for the ASD and and ASD + ADHD groups is also reported where available. The LEOPs dataset has provided data in prior publications investigating the ERG as a marker for ASD \cite{manjur2025detecting, posada2023autism, constable2024spectral, chistiakov2025time, kulyabin2024attention, albasu2024electroretinogram, kulyabin2025synthetic, kulyabin2024generating, brabec2026contour, brabec2025remodeling, brabec2023group, constable2022discrete, manjur2023spectral, constable2021photopic, kulyabin2023enhancing, npv7-8063-24, constable2020light}.

\section*{Methods}
Detailed descriptions of data collection methods and group characteristics have been previously reported \cite{constable2024spectral, lee2022electroretinogram, constable2020light}. All stimulus frequencies were at $1.96294$~$Hz$ and the date and time of day for each recording is provided along with site, sex, age, iris color as reported by the RETeval based on grey scale ratio of the pupil to a $3$~$mm$ strip on the iris at the 3 and 9 o'clock position from the pupil margin. The OPs waveforms are provided for most waveforms and were extracted \textit{post-hoc} from the averaged ERG waveform through band-pass digital filtering between $85$--$195$~$Hz$. 
(1)	The 9-step which consists of 9 randomized flash strengths recorded on a white $40$~$cd.m^{-2}$ background with 60 averages to generate the reported averaged ERG waveform. Flash strengths included $12$, $21$, $35$, $70$, $113$, $178$, $251$, $356$ and $446$~$Td.s$ on a 1130 $Td.s$ ($40$~$cd.m^{-2}$) white background with 60 averages.  The flash strengths have an equivalent $\log$~$cd.m^{-2}$ values of $-0.37$, $-0.12$, $0.11$, $0.40$, $0.60$, $0.80$, $0.95$, $1.11$ and $1.20$ based on a $6$~$mm$ pupil diameter. Note waveforms have a baseline of $20$~$ms$ and recording interval $0$--$100$~$ms$.
(2)	The 2-step protocol used only two flash strengths ($113$~$Td.s$ and $446$~$Td.s$) on a $40$~$cd.m^{-2}$ white background with 30 averages were employed to generate the reported average ERG waveform. Note the pre-stimulus baseline for these recordings is extended to  $50$~$ms$ and the recording interval is longer and from $0$ to $170$~$ms$.
(3)	The LA3 ($85$~$Td.s$) protocol was run as part of the 9-step protocol randomized series with ERG waveforms recorded at the end of the series. For the LA3 recordings were made on a on a $30$~$cd.m^{-2}$ (848 $Td.s$) white background with 30 averages. Additional Control waveforms have been added from the reference dataset at site 1.  

Clinical psychological test data for the ASD and ASD + ADHD groups are provided, including the Full scale IQ, Autism severity score based on the methods of Gotham et al., (2009) \cite{gotham2009standardizing} and Childhood Autism Rating Scale (CARS) \cite{schopler2010childhood}. All ASD or ASD + ADHD participants met diagnostic criteria according to DSM-IV-TR \cite{american2022diagnostic} or DSM-V \cite{edition2013diagnostic} which was supported either by clinical report including Autism Diagnostic Observational Schedule-2 (ADOS) \cite{lord2012} and classed as high diagnostic confidence provided by hospital based assessment; or by community based assessment supported by CARS \cite{schopler2010childhood} – classed as medium confidence or without CARS - classed as low confidence. Participants with co-occurring ASD and ADHD were included in the dataset for completeness.

\subsection*{Ethics}
The studies were conducted in accordance with the Declaration of Helsinki at sites based in Australia (Flinders University) and the United Kingdom (University College London). Ethics approvals were provided for the studies from the Southern Adelaide Clinical Human Research Ethics Committee (Approval Code: 318.16) and the Flinders University Human Research and Ethics Committee (Approval Codes: 4606 and 7180) and the Southeast Scotland Research Ethics Committee in the UK (Approval Code:18/SS/0008).

\section*{Data records}

The LEOPs dataset \cite{constable2026LEOPs} comprises 5309 ERG averaged waveforms from 253 participants (157 TD, 75 ASD, 21 ASD + ADHD), provided in three complementary formats: an Excel spreadsheet, structured JSON files, and eye image files. Data were collected across two sites (Flinders University, Australia and University College London, UK).

\subsection*{Excel spreadsheet}
The file \texttt{LEOPs\_dataset.xlsx} contains five worksheets. Three waveform sheets (\texttt{9\_step}, \texttt{2\_step}, and \texttt{LA3}) store, for each waveform, the per-recording metadata and the reported averaged ERG and, where available, OPs waveform time--amplitude signals in paired columns ($ms$, $\mu V$). The \texttt{9\_step} sheet contains 4,246 waveforms from 173 participants with 235 data points per waveform ($20$~$ms$ baseline, $0$--$100$~$ms$ recording interval). The \texttt{2\_step} sheet contains 415 waveforms from 61 participants with 430 data points per waveform ($50$~$ms$ baseline, $0$--$170$~$ms$ recording interval). The \texttt{LA3} sheet contains 648 waveforms from 217 participants with 235 data points per waveform. Two participant sheets (\texttt{Participants 9 and 2} and \texttt{participants LA3}) provide a tabular view of the per-waveform metadata including demographics, stimulus parameters, and time-domain features. Each waveform is identified by a unique waveform identifier (e.g.\ \texttt{a100-1}) comprising the participant identifier and a sequence number.

Per-waveform metadata includes: participant group (Control=~0, ASD or ASD + ADHD~=~1); diagnosis confidence (1~=~high, 2~=~medium, 3~=~low); recording site; age at test (years); sex at birth (male~=~0, female~=~1); electrode vertical position (+1, 0, $-$1, $-$2 relative to the manufacturer's recommended height of 2 mm below the lower lid); medication status; Full Scale IQ; ASD severity score \cite{gotham2009standardizing}; CARS score \cite{schopler2010childhood}; iris color ratio; test date and time; tested eye; and flash strength in both $Td.s$ and $\log$~$cd.s.m^{-2}$. Time-domain ERG features provided are the a-wave time to peak and amplitude, the b-wave time to peak  and amplitude (measured from the a-wave minimum), and the summed OPs amplitude and time where available. Of the 5309 waveforms, 4,434 (83.5\%) include OPs waveforms extracted through band-pass digital filtering ($85$--$195$~$Hz$).

\subsection*{JSON files}
To facilitate machine learning and deep learning workflows, the LEOPs dataset includes structured JSON files in the \texttt{database/jsons/} directory, with one file per participant (253 files total). Each JSON file aggregates all recordings for a participant across all protocols. The structure of each JSON file is as follows:

\noindent
\begin{minipage}[t]{0.32\textwidth}
\begin{Verbatim}[fontsize=\scriptsize,breaklines=true,breakanywhere=true]
{ "participant_id": string,
  "demographics": {
    "category": int,
    "group": string,
    "diagnosis_strength": int|null,
    "site": int,
    "sex": int,
    "medication": int,
    "fsiq": int|null,
    "asd_severity": int|null,
    "cars": float|null
  },
  "eye_images": {
    "right": string,
    "left": string
  }|null,
\end{Verbatim}
\end{minipage}\hfill
\begin{minipage}[t]{0.32\textwidth}
\begin{Verbatim}[fontsize=\scriptsize,breaklines=true,breakanywhere=true]
  "recordings": [{
    "wave_id": string,
    "protocol": string,
    "age": float,
    "notes": string|null,
    "test_date": string,
    "test_time": string,
    "test_eye": string,
    "electrode_position": int,
    "iris": float,
    "stimulus": {
      "flash_tds": float,
      "flash_cd": float,
      "frequency_hz": float
    },
\end{Verbatim}
\end{minipage}\hfill
\begin{minipage}[t]{0.32\textwidth}
\begin{Verbatim}[fontsize=\scriptsize,breaklines=true,breakanywhere=true]
    "features": {
      "a_time_ms": float,
      "a_amp_uv": float,
      "b_time_ms": float,
      "b_amp_uv": float,
      "op_sum_amp_uv": float|null,
      "op_sum_time_ms": float|null
    },
    "erg_waveform": {
      "time_ms": array[float],
      "amplitude_uv": array[float]
    },
    "op_waveform": {
      "time_ms": array[float],
      "amplitude_uv": array[float]
    }|null }]
}
\end{Verbatim}
\end{minipage}

\begin{figure}
    \centering
    \includegraphics[width=1.0\textwidth]{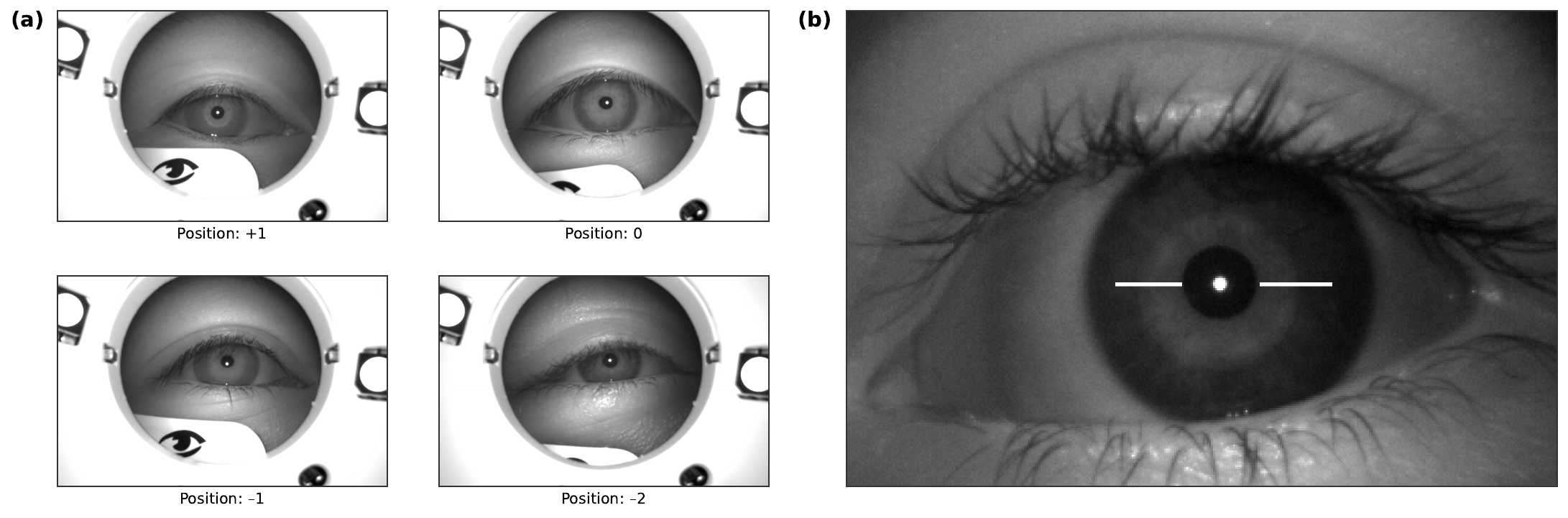}
    \caption{(a) Eye image files from the RETeval are provided as part of the dataset. Electrode position is a key factor in determining the amplitude of the recorded signal. The manufacturer recommends to place the electrode $2$~$mm$ below the lower lid. Images have been subjectively graded as +1 = $1$~$mm$ above the recommended height; 0 = at the recommended height; $-$1 or $-$2 $\sim$ $1$ or $2$~$mm$ below the recommended height respectively. (b) Iris color is calculated by the grey scale ratio of a $3$~$mm$ strip besides the pupil indicated by the white lines to the black pupil.}
    \label{figure_1}
\end{figure}

\noindent The demographics object contains participant-level fields that are constant across all recordings: numeric category (Control=~0, ASD or ASD + ADHD~=~1), diagnostic group (Control, ASD, or ASD + ADHD), diagnosis confidence (1~=~high, 2~=~medium, 3~=~low, or null if unavailable), recording site, sex (0~=~male, 1~=~female), medication status, and where available clinical scores (FSIQ, ASD severity, CARS). The \texttt{eye\_images} object links to the right and left eye image filenames, or is \texttt{null} if unavailable. The \texttt{recordings} array contains one entry per waveform, each with recording-level metadata that may vary across sessions: age (years), notes, test date and time, tested eye, electrode position (+1, 0, $-$1, $-$2 relative to the manufacturer's recommended height, graded per eye from the corresponding eye image), iris color, stimulus parameters (flash strength in $Td.s$ and $\log$~$cd.s.m^{-2}$, frequency in $Hz$), time-domain features (a-wave and b-wave timing and amplitude, summed OPs), and the raw ERG and OPs waveforms as paired time--amplitude arrays. The \texttt{op\_waveform} field is \texttt{null} for recordings without oscillatory potential data. Conversion from $Td.s$ to $cd.s.m^{-2}$ was performed assuming a $6$~$mm$ pupil diameter.

\subsection*{Eye images}
A total of 558 PNG eye images (278 right eye, 279 left eye) are provided in the \texttt{ImagesERG/} directory, covering 244 of 253 participants. Each image shows the position of the RETeval skin electrode relative to the lower eyelid and has been subjectively graded for electrode position as shown in Fig.~\ref{figure_1}. Image filenames follow the convention \texttt{\{participant\_id\}\_\allowbreak\{RightEye|LeftEye\}.png} and are cross-referenced in both the Excel and JSON files.

\begin{table}[H]
\centering
\caption{Summary of dataset parameters as mean $\pm$ SEM for each group and data set series. OPs~=~Oscillatory Potentials; TD~=~Typically Developing.}
\label{table_1}
\small
\resizebox{\textwidth}{!}{%
\begin{tabular}{l c c c c c c c c c c}
\toprule
\multicolumn{11}{l}{\textbf{Nine and Two Step protocols}} \\
\midrule
 & $N$ & Age & Sex & Iris & $a$-time & $a$-amp & $b$-time & $b$-amp & $\Sigma$OPs-t & $\Sigma$OPs-a \\
 &  & (yr) & M\,:\,F \% &  & ($ms$) & ($\mu V$) & ($ms$) & ($\mu V$) & ($ms$) & ($\mu V$) \\
\midrule
ASD & 1730 & $13.5 \pm 0.1$ & $80\,:\,20$ & $1.235 \pm 0.003$ & $12.11 \pm 0.04$ & $-6.37 \pm 0.08$ & $27.23 \pm 0.08$ & $24.32 \pm 0.26$ & $132.05 \pm 0.64$ & $22.80 \pm 0.29$ \\
ASD+ADHD & 431 & $14.5 \pm 0.2$ & $68\,:\,32$ & $1.182 \pm 0.004$ & $11.96 \pm 0.09$ & $-7.37 \pm 0.15$ & $27.27 \pm 0.15$ & $28.20 \pm 0.55$ & $132.07 \pm 1.14$ & $27.71 \pm 0.57$ \\
TD & 2500 & $13.5 \pm 0.1$ & $48\,:\,52$ & $1.258 \pm 0.003$ & $12.07 \pm 0.03$ & $-7.33 \pm 0.07$ & $27.00 \pm 0.06$ & $27.86 \pm 0.22$ & $132.84 \pm 0.48$ & $26.29 \pm 0.22$ \\
\bottomrule
\end{tabular}}
\vspace{6pt}

\resizebox{\textwidth}{!}{%
\begin{tabular}{l c c c c c c c c c c}
\toprule
\multicolumn{11}{l}{\textbf{LA3 ISCEV Standard}} \\
\midrule
 & $N$ & Age & Sex & Iris & $a$-time & $a$-amp & $b$-time & $b$-amp & $\Sigma$OPs-t & $\Sigma$OPs-a \\
 &  & (yr) & M\,:\,F \% &  & ($ms$) & ($\mu V$) & ($ms$) & ($\mu V$) & ($ms$) & ($\mu V$) \\
\midrule
ASD & 176 & $13.7 \pm 0.3$ & $80\,:\,20$ & $1.241 \pm 0.008$ & $11.48 \pm 0.08$ & $-6.80 \pm 0.22$ & $28.44 \pm 0.09$ & $28.93 \pm 0.77$ & $131.07 \pm 2.22$ & $27.13 \pm 0.83$ \\
ASD+ADHD & 37 & $15.2 \pm 0.7$ & $62\,:\,38$ & $1.196 \pm 0.016$ & $12.07 \pm 0.35$ & $-7.77 \pm 0.40$ & $27.94 \pm 0.17$ & $34.69 \pm 2.05$ & $125.07 \pm 3.54$ & $31.91 \pm 2.14$ \\
TD & 435 & $13.5 \pm 0.3$ & $37\,:\,63$ & $1.232 \pm 0.006$ & $11.58 \pm 0.05$ & $-7.17 \pm 0.11$ & $27.91 \pm 0.04$ & $32.30 \pm 0.48$ & $131.44 \pm 1.69$ & $27.37 \pm 0.60$ \\
\bottomrule
\end{tabular}}
\end{table}

\subsection*{Data overview}
Table~\ref{table_1} summarizes the LEOPs dataset  parameters for each group, presented as two sub-tables: one for the combined 9-step and 2-step protocols and one for the LA3 ISCEV standard protocol, which was collected separately and includes additional Control participants from the Flinders University reference dataset. The total number of waveforms ($N$) includes replicates within eyes across all flash strengths and excludes OPs waveforms. Mean $\pm$ SEM values are provided for age, iris color ratio, a-wave and b-wave time-domain parameters, and summed OPs amplitudes and implicit times.

The sex ratio differs markedly between groups: the ASD group shows a strong male predominance (80\,:\,20 M\,:\,F), consistent with the higher prevalence of ASD diagnoses in males \cite{american2022diagnostic}, whereas the Control group has a more balanced distribution (49\,:\,51 in the 9/2-step data). Iris color is reported as the grey-scale ratio of iris to pupil reflectance measured by the RETeval; values ~1.1 indicate lighter irides and higher values ~1.5 indicate darker irides (Fig.~\ref{figure_1}). The ASD group shows a reduced mean b-wave amplitude ($24.32 \pm 0.26$~$\mu V$) compared to TD ($27.86 \pm 0.22$~$\mu V$) across the 9-step and 2-step protocols, consistent with previous reports of attenuated retinal responses in ASD \cite{constable2020light}. The ASD + ADHD group shows intermediate or comparable values to the Control group, though with wider SEM reflecting the smaller sample size ($n$=431 ASD + ADHD vs 2500 for Control).

\begin{figure}[H]
    \centering
    \includegraphics[width=0.95\textwidth]{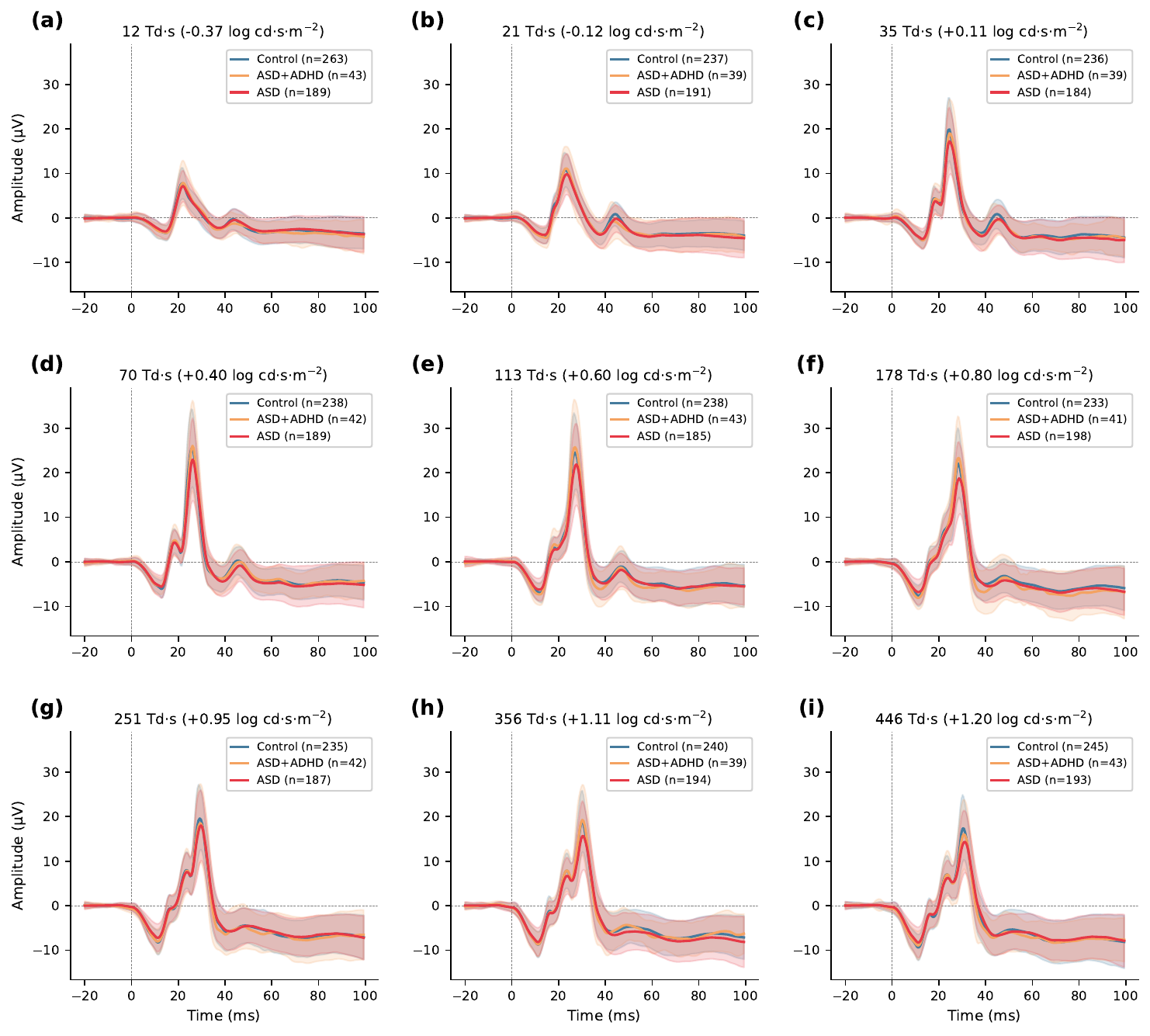}
    \caption{Mean $\pm$ 1~SD light-adapted ERG waveforms for the 9-step protocol across three groups: Control (blue), ASD (red), and ASD + ADHD (orange). Each panel shows the grand-average waveform (solid line) with shaded $\pm$1~SD region for a single flash strength. (a)~$12$~$Td.s$. (b)~$21$~$Td.s$. (c)~$35$~$Td.s$. (d)~$70$~$Td.s$. (e)~$113$~$Td.s$. (f)~$178$~$Td.s$. (g)~$251$~$Td.s$. (h)~$356$~$Td.s$. (i)~$446$~$Td.s$. The dashed vertical line indicates flash onset ($0$~$ms$); the $20$~$ms$ pre-stimulus baseline precedes it. The $n$ values denote the number of individual waveforms (including both eyes where available) contributing to each average.}
    \label{fig:9step}
\end{figure}

Fig.~\ref{fig:9step} presents the mean ERG waveforms with $\pm$1~SD shading for the 9-step protocol across all nine flash strengths, grouped by category (Control, ASD, ASD + ADHD). The characteristic photopic hill is evident as b-wave amplitude increases from $12$ to $70$~$Td.s$, plateaus around $113$--$178$~$Td.s$, and then decreases at $251$--$446$~$Td.s$. The ASD group shows a reduced b-wave amplitude relative to the Control group  at most flash strengths, consistent with prior findings \cite{constable2020light}. The ASD + ADHD group follows a similar pattern with greater variability reflecting the smaller sample size.

\begin{figure}[H]
    \centering
    \includegraphics[width=0.95\textwidth]{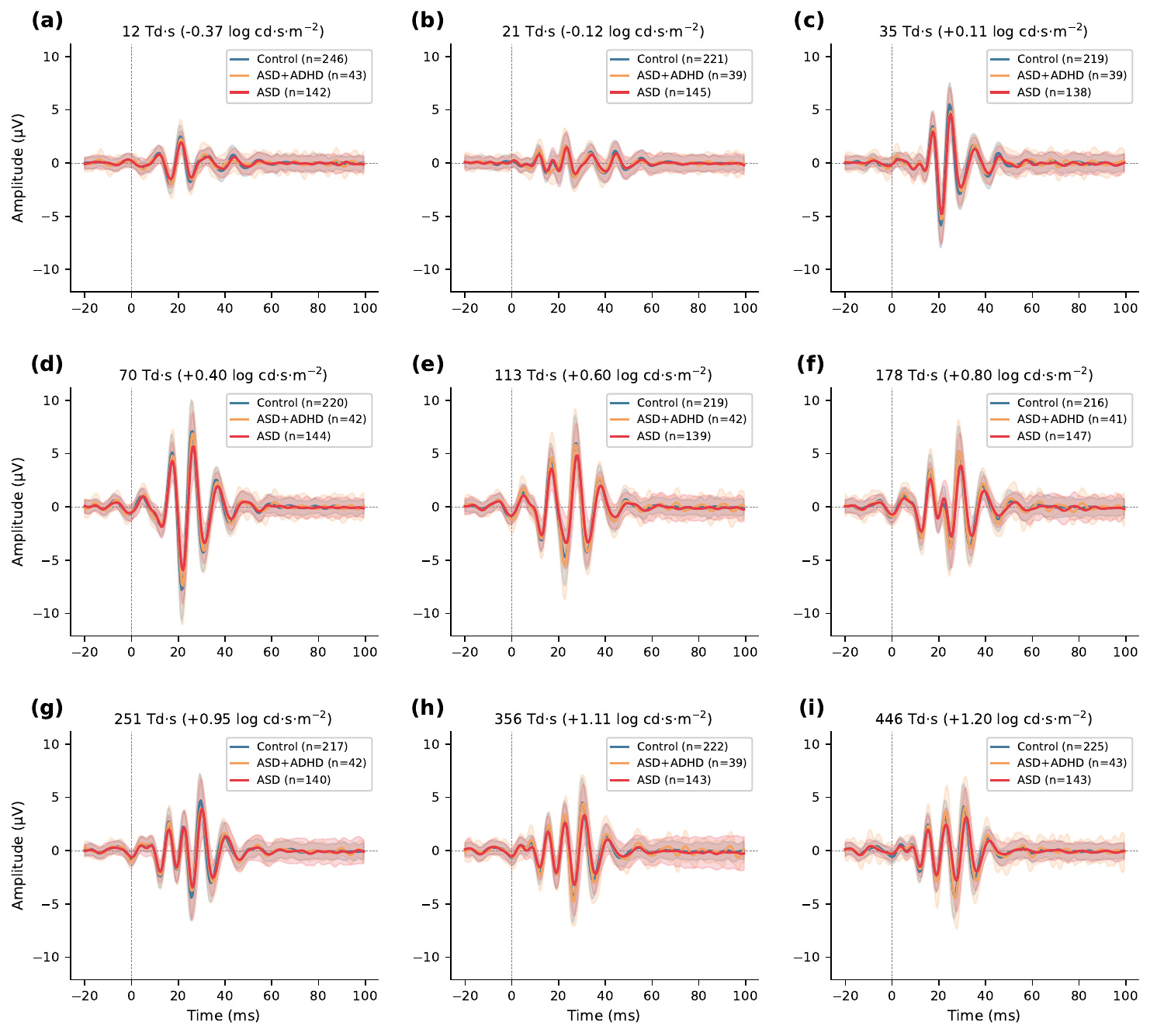}
    \caption{Mean $\pm$ 1~SD Oscillatory Potentials (OPs) waveforms for the 9-step protocol across three groups: Control (blue), ASD (red), and ASD + ADHD (orange). OPs were extracted from the averaged ERG waveform by band-pass digital filtering ($85$--$195$~$Hz$). Each panel shows the grand-average OPs waveform (solid line) with shaded $\pm$1~SD region for a single flash strength. (a)~$12$~$Td.s$. (b)~$21$~$Td.s$. (c)~$35$~$Td.s$. (d)~$70$~$Td.s$. (e)~$113$~$Td.s$. (f)~$178$~$Td.s$. (g)~$251$~$Td.s$. (h)~$356$~$Td.s$. (i)~$446$~$Td.s$. The $n$ values denote the number of individual OPs waveforms contributing to each average.}
    \label{fig:ops_9step}
\end{figure}

Fig.~\ref{fig:ops_9step} presents the mean OPs waveforms for the 9-step protocol. The OPs represent high-frequency components of the ERG waveform generated primarily by amacrine cell interactions in the inner retina. The OPs amplitude increases with flash strength across the photopic hill, with the most prominent oscillations visible at the higher flash strengths ($113$--$446$~$Td.s$).

\begin{figure}[H]
    \centering
    \includegraphics[width=0.8\textwidth]{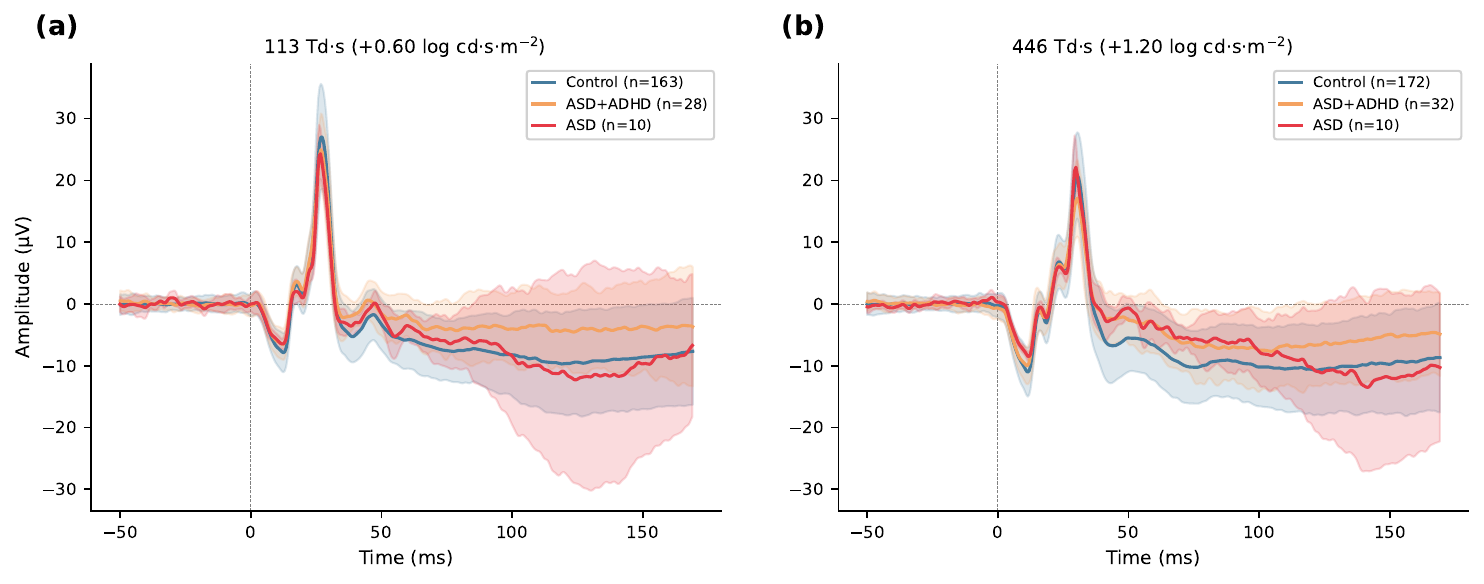}
    \caption{Mean $\pm$ 1~SD light-adapted ERG waveforms for the 2-step protocol across three groups: Control (blue), ASD (red), and ASD + ADHD (orange). (a)~$113$~$Td.s$. (b)~$446$~$Td.s$. The recording window extends from $-50$ to $170$~$ms$, capturing later waveform morphology beyond the b-wave. Oscillatory Potentials waveforms are also shown for the three groups. The $n$ values denote the number of individual waveforms contributing to each average.}
    \label{fig:2step}
\end{figure}

Fig.~\ref{fig:2step} shows the mean ERG waveforms for the 2-step protocol, which employed a longer recording window ($50$~$ms$ baseline, $0$--$170$~$ms$ post-stimulus) with two flash strengths and 30 averages. The ASD group has a notably small sample size ($n$=10 waveforms) in this protocol subset, resulting in wider SD bands. OPs are also illustrated for this dataset, with OPs also available for the 9-step protocol. 

\begin{figure}[H]
    \centering
    \includegraphics[width=0.8\textwidth]{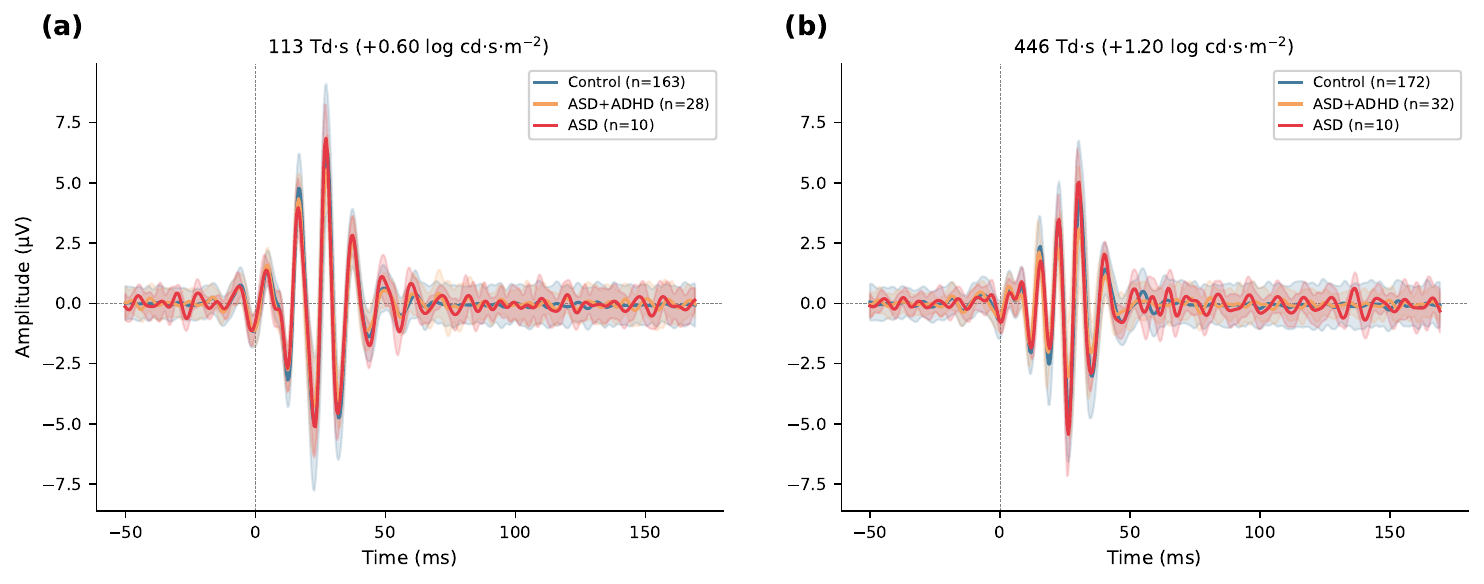}
    \caption{Mean $\pm$ 1~SD Oscillatory Potentials (OPs) waveforms for the 2-step protocol across three groups: Control (blue), ASD (red), and ASD + ADHD (orange). (a)~$113$~$Td.s$. (b)~$446$~$Td.s$. The extended recording window ($50$~$ms$ baseline, $0$--$170$~$ms$). The $n$ values denote the number of individual OPs waveforms contributing to each average.}
    \label{fig:ops_2step}
\end{figure}

\begin{figure}[H]
    \centering
    \includegraphics[width=0.7\textwidth]{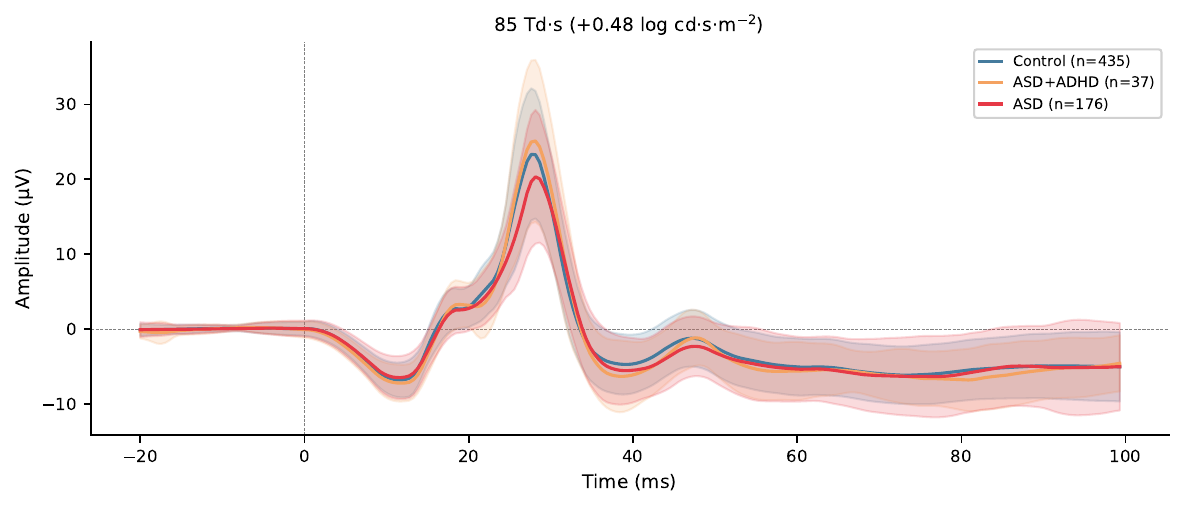}
    \caption{Mean $\pm$ 1~SD light-adapted ERG waveform for the ISCEV standard LA3 protocol ($85$~$Td.s$) across three groups: Control (blue, $n$=435), ASD (red, $n$=176), and ASD + ADHD (orange, $n$=37). The recording parameters include a  ($20$~$ms$ baseline, $0$--$100$~$ms$ recording interval, with 30 averages).}
    \label{fig:LA3}
\end{figure}

Fig.~\ref{fig:ops_2step} shows the corresponding OPs waveforms for the 2-step protocol.  The Control group ($n$=335) shows the clearest OPs morphology, while the ASD group ($n$=20) has wider variability due to the smaller sample size.

Fig.~\ref{fig:LA3} presents the ISCEV standard LA3 ($85$~$Td.s$) ERG waveform. This protocol subset includes additional Control waveforms in young adults from the Flinders University reference dataset, yielding the largest sample size in the dataset.

\section*{Technical Validation}

Signal quality was assured at two levels. First, during acquisition the RETeval device applied automatic artifact rejection: individual waveform traces containing blinks or other artifacts that fell within the upper or lower quartile of the running average were excluded before computing the reported averaged ERG waveform. Each averaged waveform therefore represents 30--60 artifact-free traces per eye. Second, a post-hoc visual inspection was performed to remove any remaining waveforms that exhibited excessive noise or large baseline offsets.

\subsection*{Data completeness.}
All 5309 ERG waveforms in the dataset contain complete time--amplitude arrays and time-domain features (a-wave and b-wave amplitude and implicit time). OPs waveforms are available for 4,434 (83.5\%) of recordings, with the remaining 875 waveforms lacking OPs data due to recording protocol differences. Eye images are provided for 244 of 253 participants (96.4\%). Eye images were excluded if the captured image occurred during a blink or eccentric fixation.

\subsection*{Demographic consistency.}
Participant-level demographic fields (diagnostic group, sex, recording site, medication status, and clinical scores) were cross-validated across all waveform sheets to verify internal consistency. Automated checks confirmed that, for each participant, these fields are identical across all recordings in the 9-step, 2-step, and LA3 protocols.

\subsection*{Stimulus--Response function.}
\begin{figure}[H]
    \centering
    \includegraphics[width=1.0\textwidth]{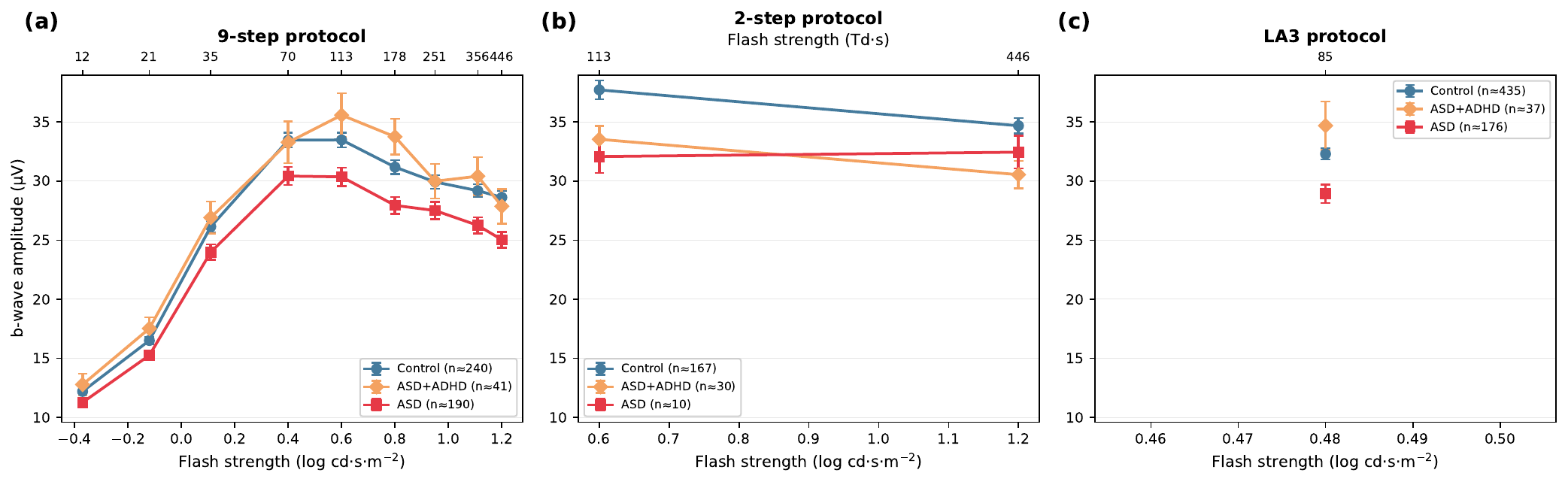}
    \caption{Stimulus--response function across all three protocols. Mean b-wave amplitude ($\pm$~SEM) is plotted against flash strength in $\log$~$cd.s.m^{-2}$ (bottom axis) and $Td.s$ (top axis) for each group: Control (blue circles), ASD (red squares), and ASD + ADHD (orange diamonds). (a)~9-step protocol: the non-monotonic photopic hill profile, with amplitude peaking near $70$--$113$~$Td.s$ before declining at higher flash strengths. (b)~2-step protocol ($113$ and $446$~$Td.s$). (c)~LA3 protocol ($85$~$Td.s$). Approximate sample sizes per group are indicated in the legend.}
    \label{fig:photopic_hill}
\end{figure}

The mean b-wave amplitude for the Control group across the 9-step protocol follows the expected photopic hill profile \cite{wali1992photopic}: amplitude increases from $12.2 \pm 4.4$~$\mu V$ at $12$~$Td.s$ to a plateau of $33.5 \pm 9.6$~$\mu V$ at $70$--$113$~$Td.s$, before declining to $28.6 \pm 8.6$~$\mu V$ at $446$~$Td.s$ (Fig.~\ref{fig:photopic_hill}a). This physiologically expected non-monotonic stimulus--response relationship confirms that the stimulus calibration and recording parameters are consistent with established norms. The 2-step protocol (Fig.~\ref{fig:photopic_hill}b) and LA3 protocol (Fig.~\ref{fig:photopic_hill}c) show consistent b-wave amplitudes at their respective flash strengths, with group differences comparable to those observed in the 9-step data.

\subsection*{Inter-ocular consistency.}
\begin{figure}[H]
    \centering
    \includegraphics[width=\textwidth]{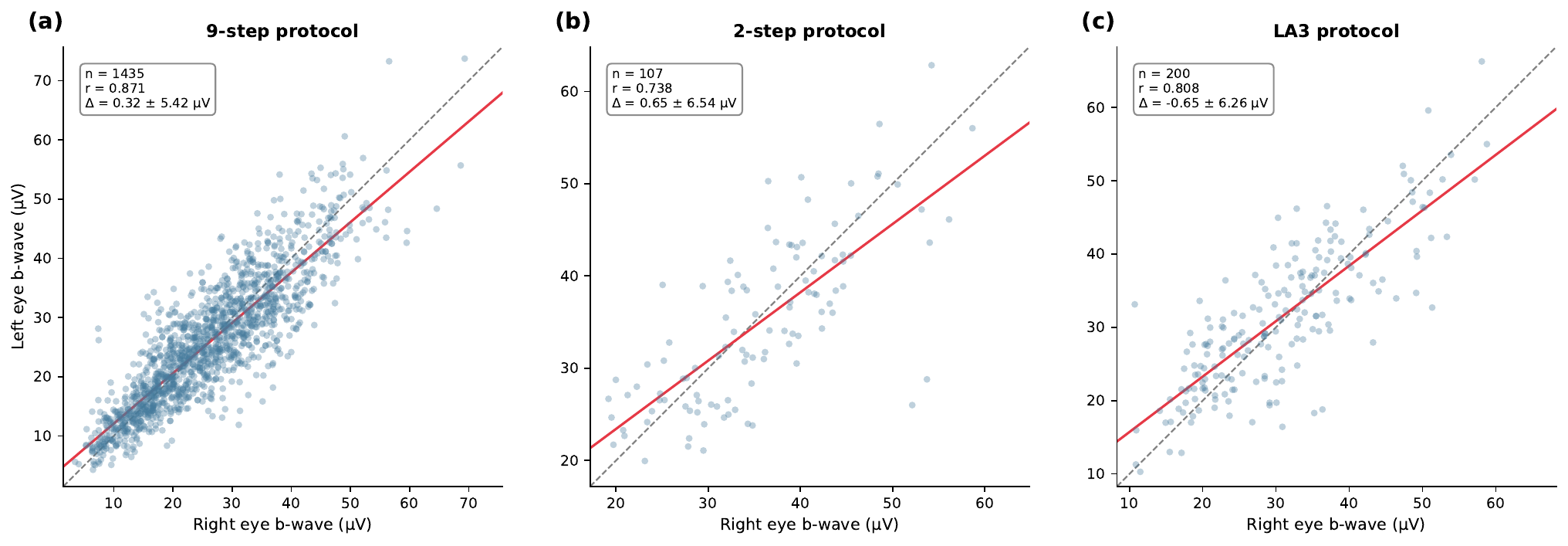}
    \caption{Inter-ocular consistency of b-wave amplitude across all three protocols. Each point represents the mean b-wave amplitude for a single participant--flash-strength combination, plotted as right eye versus left eye. The dashed grey line indicates perfect agreement (identity line) and the red line shows the linear regression fit. Pearson correlation ($r$), number of paired observations ($n$), and mean right--left difference ($\Delta$) are annotated. (a)~9-step protocol. (b)~2-step protocol. (c)~LA3 protocol.}
    \label{fig:interocular}
\end{figure}

For participants with bilateral recordings, the b-wave amplitude showed strong inter-ocular correlation across all protocols and flash strengths (Fig.~\ref{fig:interocular}). In the 9-step protocol (Fig.~\ref{fig:interocular}a), Pearson $r$ ranged from $0.67$ to $0.85$ (all $p < 10^{-20}$). Paired $t$-tests revealed no significant difference between right and left eyes at any flash strength (all $p > 0.05$), with mean inter-ocular differences of less than $1$~$\mu V$. Similar inter-ocular agreement was observed for the 2-step (Fig.~\ref{fig:interocular}b) and LA3 (Fig.~\ref{fig:interocular}c) protocols. This demonstrates good within-subject test--retest reliability of the recording setup across all stimulus conditions.

\subsection*{Cross-site comparison.}
Data were collected at two sites: Flinders University, Adelaide, Australia (Site~1; $n$ = 170 participants) and University College London, United Kingdom (Site~2; $n$ = 82 participants). Within the Control group, b-wave amplitudes were significantly higher at Site~2 than Site~1 across all flash strengths ($p < 0.005$). For example, at $113$~$Td.s$ the mean b-wave amplitude was $29.9 \pm 8.4$~$\mu V$ at Site~1 versus $35.5 \pm 9.7$~$\mu V$ at Site~2. These inter-site amplitude differences are consistent with known sources of variability in skin electrode ERG recordings, with the RETeval including differences in electrode position \cite{hobby2018effect}, pigmentation \cite{al2010light}, fixation and waveform variability \cite{arias2026mydriasis, you2023comparing}  and operator variability \cite{porciatti2004normative}. Recording site is provided as a covariate in the dataset to enable users to account for this variability in downstream analyses.

\section*{Data availability}

The LEOPs dataset available at \href{https://data.mendeley.com/datasets/w3yx7hdds7}{https://data.mendeley.com/datasets/w3yx7hdds7}.

\section*{Code availability}

The code for loading, visualizing, and validating the LEOPs dataset and used to generate the results in this paper is available at \href{https://github.com/MikhailKulyabin/LEOPs}{https://github.com/MikhailKulyabin/LEOPs}.

\bibliography{references}

\section*{Author contributions statement}

Data collection, P.C, I.L, L.L. and D.T. conceptualization, P.C and M.K. software, M.K, A.M. and A.Z. writing-original draft preparation, P.C., and M.K  writing---review and editing, P.C., M.K., D.T and A.M supervision, P.C, D.T. and A.M. All authors have read and agreed to the published version of the manuscript.

\section*{Competing interests}

The authors declare that they have no known competing financial interests or personal relationships that could have appeared to influence the work reported in this paper.

\section*{Acknowledgements}
We thank Quentin Davis and  Joshua Santosa of LKC Technologies for the custom RETeval protocols. The authors acknowledge Professor Emeritus Edward Ritvo for his inspiration and support for investigating the electroretinogram in ASD. We would also thank Professor David Skuse for his support in establishing the study in the UK.

\end{document}